\documentclass[conference]{IEEEtran}
\usepackage{mathpazo}
\usepackage{times}

\usepackage{amsmath}
\usepackage{amsfonts}
\usepackage{latexsym}
\usepackage{amssymb}
\usepackage{upref}
\usepackage{theorem}
\usepackage{graphicx}
\usepackage{psfrag}
\usepackage{cite}
\usepackage{epsfig}
\usepackage{color}
\usepackage{graphicx}

\definecolor{gray}{gray}{0.8}

\theoremstyle{plain}
\theorembodyfont{\normalfont\slshape}

\newtheorem{thm}{Theorem$\!$}

\newtheorem{clm}[thm]{Claim$\!$}

\newtheorem{lem}[thm]{Lemma$\!$}

\newtheorem{prop}[thm]{Proposition$\!$}

\newtheorem{cor}[thm]{Corollary$\!$}

\newtheorem{defn}[thm]{Definition$\!$}

\newtheorem{xmpl}[thm]{Example$\!$}

\newtheorem{cnstr}{Construction$\!$}

\newcounter{enumrom}
\renewcommand{\theenumrom}{(\roman{enumrom})}

\makeatletter
\renewcommand{\@endtheorem}{\endtrivlist}
\makeatother

\makeatletter
\renewcommand{\thefigure}{{\@arabic\c@figure}}
\renewcommand{\fnum@figure}{{\bf Figure\,\thefigure}}
\makeatother

\begin{document}


\title{\textbf{On Codes for Optimal Rebuilding Access}
\vspace*{-0.2ex}}

\author{\IEEEauthorblockN{Zhiying Wang\IEEEauthorrefmark{1}, Itzhak Tamo\IEEEauthorrefmark{1}\IEEEauthorrefmark{2}, and Jehoshua Bruck\IEEEauthorrefmark{1}\\}
\IEEEauthorblockA{\IEEEauthorrefmark{1}Electrical Engineering Department,
California Institute of Technology,
Pasadena, CA 91125, USA \\}
\IEEEauthorblockA{\IEEEauthorrefmark{2}Electrical and Computer Engineering,
Ben-Gurion University of the Negev,
Beer Sheva 84105, Israel\\}
{\it \{zhiying, tamo, bruck\}@caltech.edu}\vspace*{-2.0ex}}

\maketitle

\begin{abstract}
MDS (maximum distance separable) array codes are widely used in storage systems due to their computationally efficient encoding and decoding procedures. An MDS code with $r$ redundancy nodes can correct any $r$ erasures by accessing (reading) all the remaining information in both the systematic nodes and the parity (redundancy) nodes. However, in practice, a single erasure is the most likely failure event; hence, a natural question is how much information do we need to access in order to rebuild a single storage node? We define the \emph{rebuilding ratio} as the fraction of remaining information accessed during the rebuilding of a single erasure. In our previous work we showed that the optimal rebuilding ratio of $1/r$ is achievable (using our newly constructed array codes) for the rebuilding of any systematic node, however, all the information needs to be accessed for the rebuilding of the parity nodes. Namely, constructing array codes with a rebuilding ratio of $1/r$ was left as an open problem. In this paper, we solve this open problem and present array codes that achieve the lower bound of $1/r$ for rebuilding any single systematic or parity node.
\end{abstract}



\section{Introduction}
MDS (maximum distance separable) array codes are a family of erasure-correcting codes used extensively as the basis for RAID storage systems. An array code consists of a 2-D array where each column can be considered as a disk. We will use the term column, node, or disk interchangeably. A code with $r$ parity (redundancy) nodes is MDS if and only if it can recover from any $r$ erasures. EVENODD \cite{Shuki-evenodd} and RDP \cite{RDP-code} are examples of MDS array codes with two redundancies. In this paper, we only consider  systematic codes, namely, the information is stored exclusively in the first $k$ nodes, and the parities are stored exclusively in the last $r$ nodes.

\begin{figure}
  \centering
  \includegraphics[width=.4\textwidth]{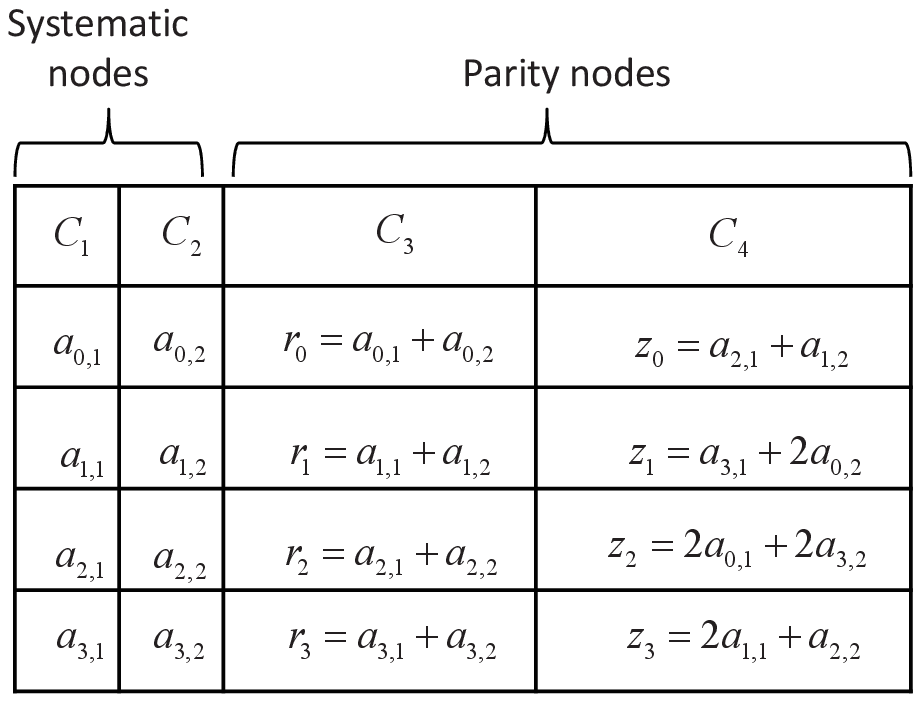}
  \caption{An MDS array code with two systematic and two parity nodes. All the elements are in finite field $F_3$. The first parity column $C_3$ is the row sum and the second parity column $C_4$ is generated by the zigzags. For example, zigzag $z_0$ contains the elements $a_{i,j}$ that satisfy $f_j^1(i)=0$.}
  \label{fig1}
\end{figure}

In order to correct $r$ erasures, it is obvious that one has to access (or read) the information in all the surviving nodes. However, in practice it is more likely to encounter a single erasure rather than $r$ erasures. So a natural questions is: How much information do we need to access when rebuilding a single erasure? Do we have to access all the surviving information? We define the \emph{rebuilding ratio} as the ratio of accessed information to the remaining information in case of a single erasure. For example, it is easy to check that for the code in Figure \ref{fig1}, if any two columns are erased, we can still recover all the information, namely, it is an MDS code. However, if column $C_1$ is erased, it can be rebuilt by accessing $a_{0,2},a_{1,2}$ from column $C_2$, $r_0,r_1$ from column $C_3$, and $z_0,z_1$ from column $C_4$, as follows:
\begin{eqnarray*}
a_{0,1} &= 2a_{0,2} + r_{0} \\
a_{1,1} &= 2a_{1,2} + r_{1} 
\end{eqnarray*}
\begin{eqnarray*}
a_{2,1} &= 2a_{1,2} + z_{0}\\ 
a_{3,1} &= a_{0,2} + z_{1}
\end{eqnarray*}
Here all elements are in finite field $F_3$. Hence, by accessing only half of the remaining information, the erased node can be rebuilt. Details on this new code will be discussed in Section \ref{sec2}.

A related problem called {\em repair bandwidth} was first proposed in \cite{Dimakis2010}. The paradigm there is that one can access the entire information and perform computations within each node, and the question is how much information is {\em transmitted} for rebuilding? A lower bound on the repair bandwidth was given in \cite{Dimakis2010}. When a single erasure occurs and all the remaining nodes are accessible, the lower bound for the bandwidth is $\frac{1}{r}$. Recently, a number of codes were designed to achieve the bandwidth lower bound. When the number of parity nodes is larger than that of the systematic nodes, explicit code constructions were given in \cite{Rashmi,Kumar2009,Suh-alignment}. For all cases, \cite{cadambe10,Suh10} achieved the lower bound asymptotically.

It is clear that a lower bound on the repair bandwidth is also a lower bound on the rebuilding ratio. In \cite{old} we presented an explicit construction of MDS array codes  that achieve the lower bound on the ratio for rebuilding any {\em systematic node}. A similar code construction was given in \cite{cadambe}. Also in \cite{Papai} a similar code with $2$ parities was proposed - it has optimal repair bandwidth for any single erasure.

The main contribution of this paper is an explicit construction of MDS array codes with $r$ parity nodes, that achieves the lower bound $1/r$ for rebuilding {\em any systematic or parity node}. The rebuilding of a single erasure has an efficient implementation as computations within nodes are not required. Moreover, our codes have simple encoding and decoding procedures - when $r=2$ and $r=3$, the codes require finite-field sizes of $3$ and $4$, respectively.

The rest of the paper is organized as follows. Section \ref{sec2} introduces the rebuilding ratio problem for MDS array codes and reviews the code construction in \cite{old}. Section \ref{sec3} describes the construction of our codes with optimal rebuilding ratio. Finally, the paper is summarized in Section \ref{sec4}.

\section{Rebuilding Ratio Problem}
\label{sec2}
In this section we formally define the rebuilding ratio problem and review the code construction in \cite{old}. We then prove that the construction can be made an MDS code, in fact, this will be the basis for proving that our newly proposed construction which is described in Section \ref{sec3} is also an MDS code.

We first define the framework of a systematic MDS array code. Let $A=(a_{i,j})$ be an information array of size $p \times q$. A column is also called a node, and an entry is called an element. Each of the $q$ columns is a systematic node in the code. We add $r$ parity columns to this array on the right, such that from any $q$ columns, we can recover the entire information. In \cite{old}, it was shown that if each information element is protected by exactly $r$ parity elements, then each parity node corresponds to $q$ permutations acting on $[0,p-1]$. More specifically, suppose the permutations are $f_1,f_2,\dots,f_q$. Then the $t$-th element in this parity node is a linear combination of all elements $a_{i,j}$ such that $f_j(i)=t$. The set of information elements contained in this linear combination is called a \emph{zigzag set}. For the $t$-th element in the $l$-th parity, $t \in [0,p-1],l \in [0,r-1]$, denote by $f_1^l,\dots,f_q^l$ the set of associated permutations, and $Z_t^l$ the zigzag set.

The ordering of the elements in each node can be arbitrary, hence, we can assume that the first parity node is always a linear combination of each row (corresponding to identity permutations). Figure \ref{fig1} is an example of such codes. The first parity $C_3$ corresponds to identity permutations. The second parity $C_4$ corresponds to the permutations
\begin{align*}
  f_1^1&=(2, 3, 0, 1), \\
  f_2^1&=(1, 0, 3, 2).
\end{align*}

For a given MDS code with parameters $q,r$, we ask what is the accessed fraction in order to rebuild a single node (in the average case)?
Hence, the {\em rebuilding ratio} of a code is:
$$R=\frac{\sum_{i=1}^{q+r}(\#\text{ accessed elements to rebuild node } i)}{(q+r)(\# \text{ remaining elements})}. $$

When a systematic node is erased, we rebuild each unknown element by one of the parity nodes. That is, we access one parity element containing the unknown, and access all the elements in the corresponding zigzag set except the unknown. In order to lower the number accesses, we would like to find (i) good permutations such that the accessed zigzag sets intersect as much as possible, and (ii) proper coefficients in the linear combinations such that the code is MDS. For example, in Figure \ref{fig1}, in order to rebuild column $C_1$, we access the zigzag sets $A=\{Z_0^0,Z_1^0\},B=\{Z_0^1,Z_1^1\}$, corresponding to parities $\{r_0,r_1\},\{z_0,z_1\}$. The surviving elements in $A$ and in $B$ are identical, i.e., $\{a_{0,2},a_{1,2}\}$, therefore, only $1/2$ of the elements are accessed. Besides, the coefficients $\{1,2\}$ in the parity linear combinations guarantee that any two nodes are sufficient to recover all the information. Hence the code is MDS.

Next we review the construction with optimal rebuilding for systematic nodes that was presented in \cite{old}. The idea in the code construction was to form permutations based on $r$-ary vectors.

Let $e_1,e_2,\dots,e_k$ be the standard vector basis of $\mathbb{Z}_r^k$. We will use $x$ to represent both an integer in $[0,r^k-1]$ and its $r$-ary expansion (the $r$-ary vector of length $k$). It will be clear from the context which meaning is used. All the calculations are done over $\mathbb{Z}_r$

\begin{cnstr}
\label{cnstr1}
Let the information array be of size $r^k \times k$.
Define permutation $f_j^l$ on $[0,r^k-1]$ as $f_j^l(x)=x+le_j$, $j\in [1,k],l \in [0,r-1]$. For $t \in [0, r^k-1]$, we define the zigzag set $Z_t^l$ in parity node $l$ as the elements $a_{i,j}$ such that their coordinates satisfy $f^l_j (i) = t$. Let $Y_j=\{x \in [0,r^k-1]: x \cdot e_j=0\}$. Rebuild column $j$ by accessing rows $Y_j$ in all remaining columns.
\end{cnstr}

\begin{thm}
Construction \ref{cnstr1} has optimal ratio $1/r$ for rebuilding any systematic node \cite{old}.
\end{thm}

Figure \ref{fig1} is an example of Construction \ref{cnstr1}. As mentioned before, only $1/2$ of the information is accessed in order to rebuild $C_1$. The accessed elements are in rows $Y_1=\{x \in [0,3]: x \cdot e_1=0\}=\{0,1\}$.

Next, we show that by assigning the coefficients in the parities properly, the code is MDS.
Let $P_j=(a_{i,l})$ be the permutation matrix corresponding to $f_j=f_j^1$, namely, $a_{i,l}=1$ if $l+e_j=i$, and $a_{i,l}=0$ otherwise. Assigning the coefficients is the same as modifying $a_{i,l}=1$ to other non-zero values. When $r=2,3$, modify $a_{i,l}=1$ to $a_{i,l}=c$, if $l \cdot \sum_{t=1}^{j} e_{t} =0$, where $c$ is an primitive element of $F_3,F_4$, respectively. The above assignment will make the code MDS for $r=2,3$ \cite{old}. For example, the coefficients in Figure \ref{fig1} is assigned in this way.

When $r \ge 4$, modify all $a_{i,l}=1$ to $a_{i,l}=\lambda_j$, for some $\lambda_j$ in a finite field $F$. Let the generator matrix of the code be
$$ G'=\left[
\begin{array}{ccc}
I & & \\
& \ddots & \\
& & I \\
I & \cdots & I \\
P_1^1 & \cdots & P_k^1 \\
\vdots & & \vdots \\
P_1^{r-1} & \cdots & P_k^{r-1}
\end{array}
\right]. $$
The following theorem shows that under this assignment the code can be MDS.

\begin{thm}
\label{thm1}
(1) Construction \ref{cnstr1} can be made an MDS code for a large enough finite field. \\
(2) When $r=2,3$, field of size $3$ and $4$ is sufficient to make the code MDS.
\end{thm}
\begin{IEEEproof}
Part (2) was given in \cite{old}. We only prove part (1).
An MDS code means that it can recover any $r$ erasures. Suppose $t$ systematic nodes and $r-t$ parity nodes are erased, $1 \le t \le r$. Thus suppose we delete from $G'$ the systematic rows $\{j_1,j_2,\dots,j_t\}$ and the remaining parity nodes are $\{i_1,i_2,\dots,i_t\}$. Then the following $t \times t$ block matrix should be invertible:
\begin{equation}
\label{eq3}
  G=\left[
  \begin{array}{ccc}
    P_{j_1}^{i_1} & \cdots & P_{j_t}^{i_1} \\
    \vdots & & \vdots \\
    P_{j_1}^{i_t} & \cdots & P_{j_t}^{i_t}
  \end{array}
  \right]
\end{equation}
Its determinant $\det(G)$ is a polynomial with indeterminates $\lambda_{j_1},\dots,\lambda_{j_t}$. All terms have highest degree $r^k(i_1+\dots+i_t)$. One term with highest degree is $\prod_{s=1}^{t}\lambda_{j_s}^{i_s r^k}$ with non-zero coefficient $1$ or $-1$. So $\det(G)$ is a non-zero polynomial. Up to now we only showed one possible case of erasures. For any $r$ erasures, we can find the corresponding non-zero polynomial. The product of all these polynomials is again a non-zero polynomial. Hence by \cite{Alon} for a large enough field there exist assignments of $\{\lambda_j\}$ such that the polynomial is not $0$. Then each $G$ is invertible, and the code is MDS.
\end{IEEEproof}

\section{Code Construction}
\label{sec3}
The code in \cite{old} has optimal rebuilding for systematic nodes. However, in order to rebuild a parity node, one has to access all the information elements.
In this section we construct MDS array codes with optimal rebuilding ratio for rebuilding both the systematic and the parity nodes. The code has $k-1$ systematic nodes and $r$ parities nodes, for any $k,r$.

 Consider the permutation $f_j=f_j^1$ in Construction \ref{cnstr1}. It is clear that $f_j$ is a permutation of order $r$, i.e., $f_i^r$ is the identity permutation. For $i \in [0,r-1]$, define $X_i$ as the set of vectors of weight $i$, namely, $X_i=\{v \in \mathbb{Z}_r^k: v \cdot (1,\dots,1)=i\}$.
$X_0$ is a subgroup of $\mathbb{Z}_r^k$ and $X_i=X_0+i e_k$ is its coset, where $e_k=(0,\dots,0,1)$. Assume the elements in $X_i$ are ordered, $i \in [0,r-1]$, and the ordering is
\begin{align*}
X_0 &= (v_1,\dots,v_{r^{k-1}}), \\
X_i &= (v_1+ie_k,\dots,v_{r^{k-1}}+ie_k).
\end{align*}
Since the ordering of the elements in each column does not matter, we can reorder them as $(X_0,X_1,\dots,X_{r-1})$, with each $X_i$ ordered as above. One can check that $f_j(X_i)=X_{i+1}$, where the subscript is added mod $r$. So the matrix $P_j$ can be written as
\begin{equation}
\label{eq1}
P_j=\bordermatrix{
~ & X_0 & X_1 & \dots & X_{r-1} \cr
X_0 & & & & p_j \cr
X_1 & p_j & & & \cr
\vdots & & \ddots & & \cr
X_{r-1} && & p_i &},
\end{equation}
where $p_j$ corresponds to the mapping of $f_j:X_i \mapsto X_{i+1}$. In particular, if $p_j$ is viewed as a permutation acting on $X_0$, then for $x \in X_0$,
$$p_j(x)=x+e_j-e_k.$$
When $r=2,3$, modify the $1$ entries of $p_i$ into $c$ if its corresponding column $l$ satisfies $l \cdot \sum_{t=1}^{j} e_t=0$. Here $c$ is  an primitive element in $F_3,F_4$. When $r \ge 4$, modify $1$ entries into $\lambda_j$.

In the following, we will use blocks the same as single elements. When referring to row or column indices, we mean block row or column indices. We refer to $p_j$ as a small block, and the corresponding block row or column as a small block row or column. And $P_j$ is called a big block with big block row or column. Moreover, we assume the elements in each column are in order $(X_0,\dots,X_{r-1})$.

\begin{figure}
  \centering
  \includegraphics[width=.4\textwidth]{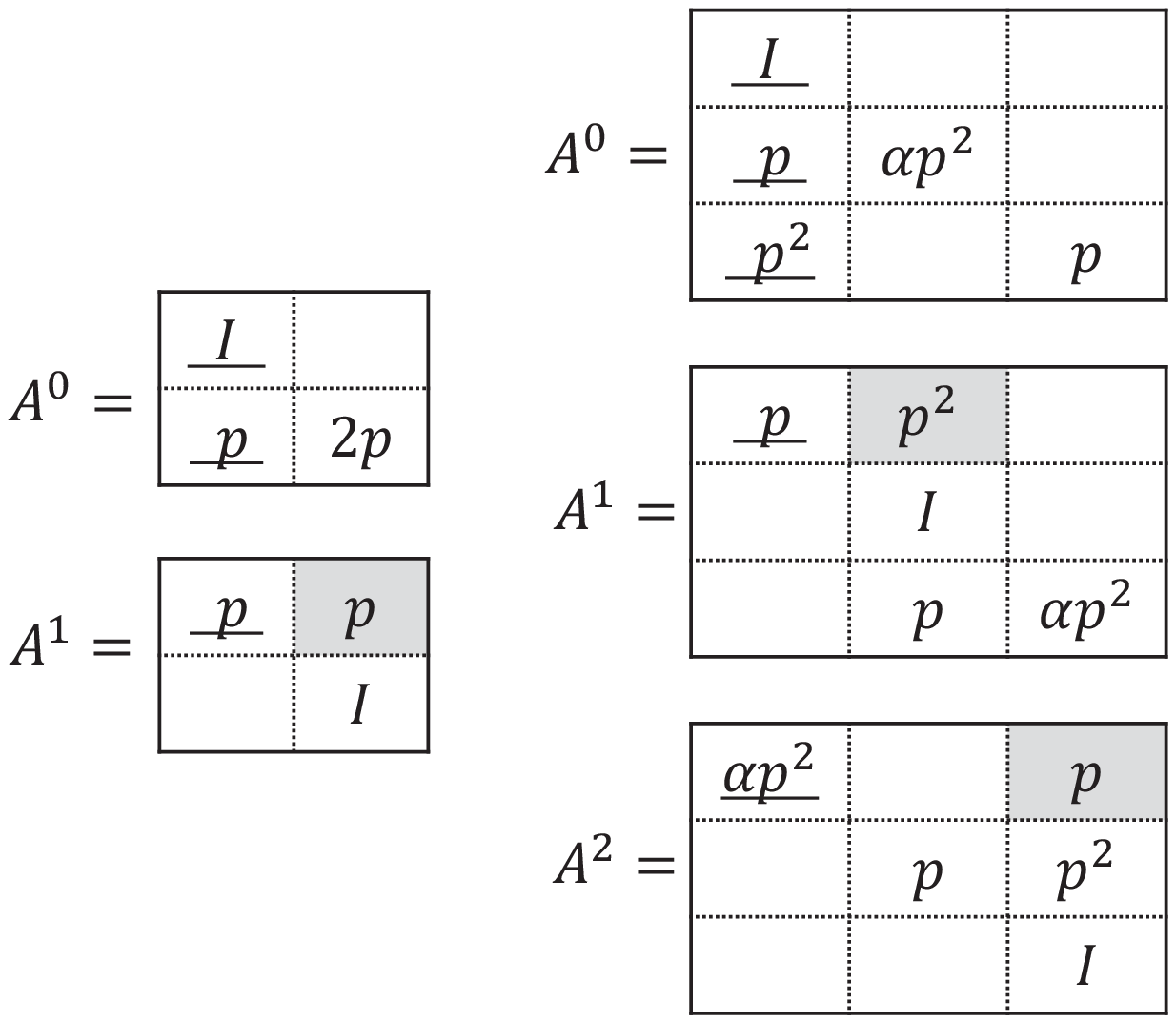}
  \caption{Parity matrices $A^i$ for $r=2$ (left) and $r=3$ (right) parities. When the first parity node is erased, the underlined elements are accessed from systematic nodes. The remaining unknown elements are recovered by the shaded elements from parity nodes.}
  \label{fig2}
\end{figure}

\begin{cnstr}
  \label{cnstr2}
  Suppose the information array is of size $r^k \times (k-1)$.
  For $j\in [1,k-1]$, define a big block matrix
  \vspace{-0.5cm}
  $$A_j^0=
  \bordermatrix{
     & & & & & &\cr
     0 & I & & & & & \cr
     1 & p_j  & \alpha p_j^{r-1}& & & &\cr
     2 & p_j^2 & & \alpha p_j^{r-2} & & &\cr
     \vdots & \vdots & & & \ddots & &\cr
     r-2 & p_j^{r-2} & & & & p_j^{2} & \cr
     r-1 & p_j^{r-1} & & & & & p_j}
  $$
  where $\alpha \neq 0,1$ is an element of the finite field and is multiplied to the diagonal in rows $1,\dots,\lfloor\frac{r}{2}\rfloor$.
  And define $A_j^i$ by cyclicly shifting the rows and columns of $A_j^0$ to the right and bottom by $i$ positions:
  $$A_j^i=\left[
  \begin{array}{cccccc}
    \beta p_j^i & & & p_j^{r-i} & & \\
    & \ddots & & \vdots & & \\
    & & p_j & p_j^{r-1} & & \\
    & & & I & & \\
    & & & p_j & \alpha p_j^{r-1} & \\
    & & & \vdots & & \ddots
  \end{array}
  \right],
  $$
  where $\beta=\alpha$ or $1$. If $x-i < \frac{r}{2}$ or $x-i=\frac{r}{2}, i < \frac{r}{2}$, coefficient $\alpha$ is multiplied to the diagonal in row $x$.
  Construct the code as follows. Let the first $k-1$ nodes be systematic, and the last $r$ nodes be parities. Parity $i$ is defined by $A_1^i,\dots,A_{k-1}^i$. The generator matrix is
  $$\left[
  \begin{array}{ccc}
    I & & \\
    & \ddots & \\
    & & I \\
    A_1^0 & \cdots & A_{k-1}^0 \\
    \vdots & & \vdots \\
    A_1^{r-1} & \cdots & A_{k-1}^{r-1}
  \end{array}
  \right].$$
\end{cnstr}

Sometimes we will omit the subscript $j$ when it is not important, and the superscript is computed mod $r$.

\begin{figure}
  \centering
  \includegraphics[width=.48\textwidth]{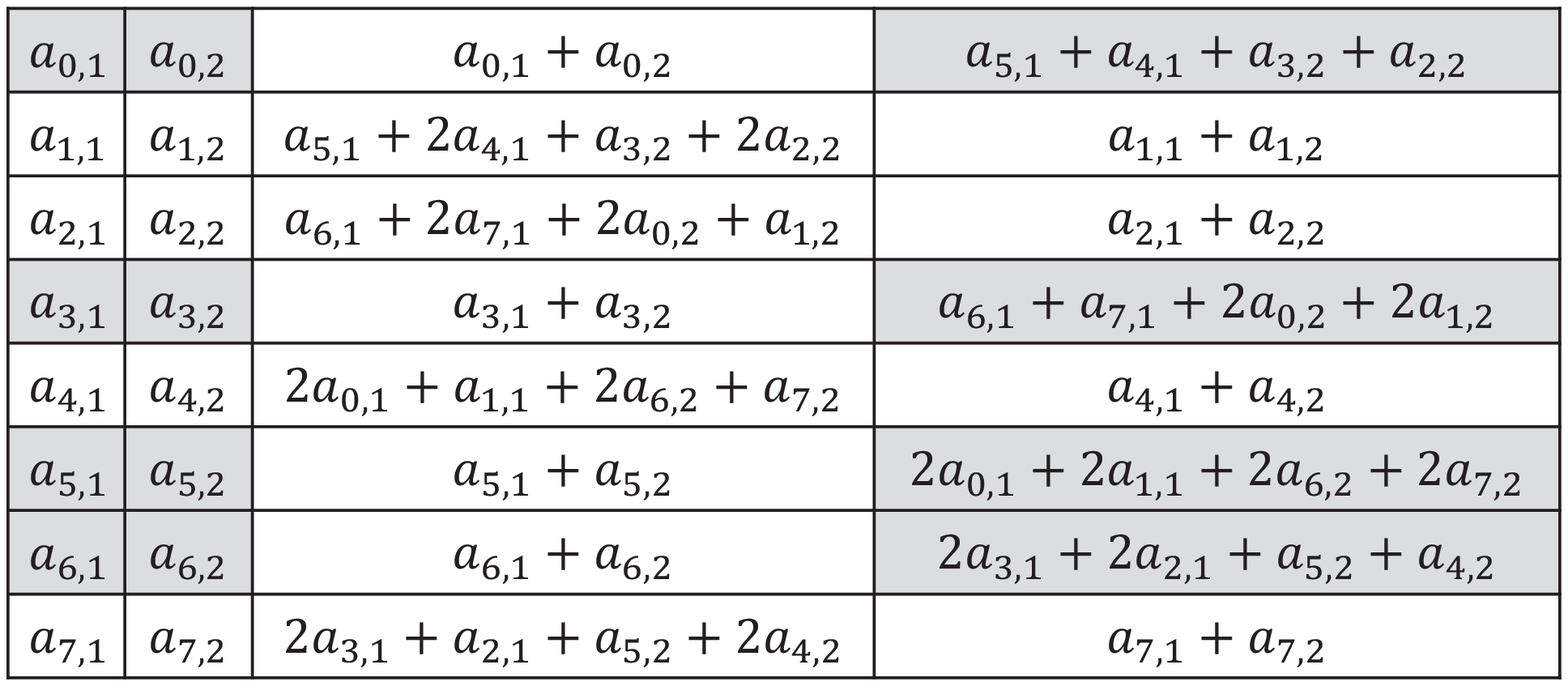}
  \caption{An MDS array code with two systematic and two parity nodes by Construction \ref{cnstr2}. The finite field used is $F_3$. The shaded elements are accessed to rebuild the first parity node.}
  \label{fig3}
\end{figure}

\begin{xmpl}
\label{xmpl1}
  For two and three parities, the matrices $A^i$ are shown in Figure \ref{fig2}. When $r=2$, as finite field $F_3$ is used, we can take $\alpha=2 \neq 1$. Coefficient $\alpha=2$ is multiplied to only the second diagonal in $A^0$.
  When $r=3$, finite field $F_4$ is used and we choose some $\alpha \neq 0,1$. We multiply $\alpha$ to one diagonal block in each $A^i$. An example of a code with $2$ parities is shown in Figure \ref{fig3}.
\end{xmpl}

Next we show that the code in Construction \ref{cnstr2} has optimal ratio.
We first observe that in $A^i$, the $x$-th row is
$$\bordermatrix{
~ & &         i        &       &   x      & \cr
~ & \cdots & p^{x-i} & \cdots & \beta p^{i-x} & \cdots },
$$
where the values above are the column indices and omitted blocks are all zero. Here $\beta=\alpha$ if $x-i < \frac{r}{2}$ or $x-i=\frac{r}{2}, i < \frac{r}{2}$, and $\beta=1$ otherwise. Therefore, suppose $i'-i < \frac{r}{2}$ or $i'-i=\frac{r}{2}, i < \frac{r}{2}$, then the $i'$-th row in $A^i$ and the $i$-th row in $A^{i'}$ are the same except for the coefficients:
\begin{equation}
\label{eq2}
\bordermatrix{
 ~  & &         i        &       &   i'      & \cr
 i' \text{ in } A^i &  \cdots & p^{i'-i} & \cdots & \alpha p^{i-i'} & \cdots \cr
 i \text{ in } A^{i'}~ & \cdots & p^{i'-i} & \cdots &  p^{i-i'} & \cdots
}.
\end{equation}

\begin{thm}
  The code has ratio $1/r$ for rebuilding any node.
\end{thm}
\begin{IEEEproof}
  {\bf Systematic rebuilding:} w.l.o.g. assume column $e_1$ is erased. Access equations $Y=\{v \in \mathbb{Z}_r^k: v \cdot e_1=0\}$ from each parity. We first show that all the unknowns $(x_0,\dots,x_{r^k-1})$ in column $e_1$ are solvable from these equations. For all $l \in Y$, $x_l$ is contained in equation
  $$x_l$$
  because of the small row block $[\cdots I \cdots].$
  Notice that $Y$ is a subgroup of $\mathbb{Z}_r^k$, and $Y-t e_k = Y$ for any $t \in [0,r-1]$.
  For any $l \in Y$, suppose $l \in Y \cap X_{i'}$ for some $i'$, so $l+(i-i')e_k \in Y \cap X_{i}$ for all $i \in [0,r-1]$.
  In \eqref{eq2} consider row $l$ in $A^i$ and row $l+(i-i')e_k$ in $A^{i'}$, and write $t=i'-i \le \lfloor \frac{r}{2} \rfloor$. Then we have equations
  \begin{eqnarray*}
  b x_{l-t e_1}+\alpha c x_{l+t(e_1-e_k)}, \\
  b x_{l-t e_1}+ c x_{l+t(e_1-e_k)},
  \end{eqnarray*}
  for some coefficients $\alpha \neq 0,1$ and $b,c \neq 0$. These equations are obviously independent.
  Moreover since $l+t(e_1-e_k) \in Y+t e_1$, we can solve unknowns indexed
  $$\cup_{t=0}^{r-1}Y+t e_1 =[0,r^k-1].$$
  Hence all unknowns are solvable.

  Next we show that the fraction of elements accessed in the remaining columns is $1/r$. For a parity node $A^i$, only rows $Y$ are accessed, which is a fraction of $1/r$. The corresponding columns in $A^i$ of theses equations are accessed from the systematic nodes.
  For a surviving systematic node $j \in [2,k-1]$ and parity $i$, by definition of $p_j^i$, rows $Y$ in $A_j^i$ are mapped to columns $Y'=Y+i(e_k-e_j)+se_k$ for some $s$. However, $Y'$ is a coset of $Y$ and since $i(e_k-e_j)+se_k \in Y$, we have $Y'=Y$. Thus only elements with indices $Y$ are accessed from each node.

  {\bf Parity rebuilding:} Since the parities are all symmetric, w.l.o.g. suppose the first parity is erased. Access $X_0$ from each node, which is the set of vectors of weight $0$. Need to show this is sufficient to recover
  $$A=[A_1^0,\dots,A_{k-1}^0],$$
  where $A_j^0$ is defined in Construction \ref{cnstr2}. Since $X_0$ is sent from the systematic nodes, the $0$-th column in each big block is known, and we can remove them from the equations. By \eqref{eq2}, from parity $i'$ we can access row
  $$[\cdots \underline{\beta' p_1^{i'}} \cdots p_1^{-i'} \cdots \underline{\beta' p_2^{i'}} \cdots p_2^{-i'} \cdots],$$
  where the underlined elements are known from the systematic nodes and can be treated as 0. Here $\beta'$ is $1$ or $\alpha$. Multiplying this row by $\beta$, we can rebuild the $i'$-th row of $A$:
  $$[\cdots \underline{p_1^{i'}} \cdots \beta p_1^{-i'} \cdots \underline{p_2^{i'}} \cdots \beta p_2^{-i'} \cdots],$$
  where $\beta \beta'=\alpha$ and $i'=1,2,\dots,r-1$. The $0$-th row is rebuilt from the systematic nodes directly. Thus the erased node is rebuilt by accessing $X_0$, which is $1/r$ of the elements.
\end{IEEEproof}

\begin{xmpl}
  \label{xmpl2}
  Consider the code with two or three parities in Figure \ref{fig2}. When the first parity node is erased, one can access $X_0$ from the systematic nodes, and the underlined elements are known. Then access the shaded elements from the surviving parity nodes. It is easy to see that the first parity can be rebuilt from the accessed elements.

  For the specific example of Figure \ref{fig3}, when the first systematic node is erased, one can access rows $0,1,2,3$ from all surviving nodes. When the first parity node is erased, one can access rows $0,3,5,6$ from all the remaining nodes (the shaded elements). Then it is easy to check that in both cases it is sufficient to rebuild the erased column.
\end{xmpl}

Next we show the construction is indeed an MDS code. We prove this by reducing this problem to the fact that Construction \ref{cnstr1} is MDS. First we make an observation on the small blocks.

\begin{lem}
\label{lem1}
  Construction \ref{cnstr1} is MDS iff any $t \times t$ sub block matrix of
  \[H'=\left[\begin{matrix}
    p_1^0 & \cdots & p_k^0 \\
    \vdots & & \vdots \\
    p_1^{r-1} & \cdots & p_k^{r-1}
  \end{matrix}\right]\]
  is invertible, for all $t \in [1,r]$.
\end{lem}
\begin{IEEEproof}
  Consider the $t \times t$ sub block matrix of $H'$:
  \[H=\left[\begin{matrix}
    p_1^0 & \cdots & p_t^0 \\
    \vdots & & \vdots \\
    p_1^{t-1} & \cdots & p_t^{t-1}
  \end{matrix}\right].\]
  We showed in Theorem \ref{thm1} that Construction \ref{cnstr1} is MDS iff any $G$ in \eqref{eq3} is invertible. W.l.o.g. suppose $\{i_1,\dots,i_t\}=\{0,\dots,t-1\}, \{j_1,\dots,j_t\}=\{1,\dots,t\}$. By \eqref{eq1}, $G$ can be rewritten as
  $$G=\left[\begin{array}{cccc|cccc|c}
    \colorbox{gray}{$I$} & & & & \colorbox{gray}{$I$} & & & & \\
    & I & & & & I & & & \dots \\
    & & \ddots & & & & \ddots & & \\
    & & & I & & & & I & \\
    \hline
    & & & p_1 & & & & p_2 & \\
    \colorbox{gray}{$p_1$} & & & & \colorbox{gray}{$p_2$} & & & & \dots \\
    & \ddots & & & & \ddots & & & \\
    & & p_1 & & & & p_2 & & \\
    \hline
    & & p_1^2 & & & & p_2^2 & & \\
    & & & p_1^2 & & & & p_2^2 & \dots \\
    \colorbox{gray}{$p_1^2$} & & & & \colorbox{gray}{$p_2^2$} & & & & \\
    & \ddots & & & & \ddots & & & \\
    \hline
    & \vdots & & & & \vdots & &
  \end{array}\right],$$
  where each big block is composed of $r \times r$ small blocks. We can see that the shaded small blocks are the only non-zero blocks in their corresponding rows and columns, and they form the sub-matrix $H$. Therefore $G$ being invertible is equivalent to $H$ and the remaining sub-matrix both being invertible. Moreover the remaining sub-matrix has a similar form as $G$ and we can again find $t$ rows and $t$ columns corresponding to $H$. Continue this we get
  $$\det(G) \neq 0 \Leftrightarrow (\det(H))^r \neq 0 \Leftrightarrow \det(H) \neq 0.$$
  The same conclusion holds for any sub matrix of $H'$. Thus completes the proof.
\end{IEEEproof}

The method of taking out sub block matrices to compute the determinant as above is also used in the proof of the following theorem, which shows that Construction \ref{cnstr2} is indeed an MDS code.

\begin{thm}
If the coefficients in the linear combinations of the parities are chosen such that Construction \ref{cnstr1} is MDS, then Construction \ref{cnstr2} is also MDS.
\end{thm}
\begin{IEEEproof}
  Similar to Theorem \ref{thm1}, Construction \ref{cnstr2} being MDS means any of the following matrix is invertible:
  $$A=\left[\begin{matrix}
    A_{j_1}^{i_1} & \cdots & A_{j_t}^{i_1} \\
    \vdots & & \vdots \\
    A_{j_1}^{i_t} & \cdots & A_{j_t}^{i_t}
  \end{matrix}\right]_{rt \times rt} ,$$
  where $t \in [1,r], I=\{i_1,\dots,i_t\} \subseteq [0,r-1], \{j_1,\dots,j_t\} \subseteq [1,k-1]$. Let the complement of $I$ be $\overline{I}=[0,r-1] \backslash I$. In each big block consider the small block column $x \in \overline{I}$. Only small block rows $x$ in each big block are non-zero. Thus we can take out this $t \times t$ sub block matrix:
  $$\left[
  \begin{matrix}
  \beta_1 p_{j_1}^{i_1-x} & \cdots & \beta_1 p_{j_t}^{i_1-x} \\
  \vdots & & \vdots \\
  \beta_t p_{j_1}^{i_t-x} & \cdots & \beta_t p_{j_t}^{i_t-x} \\
  \end{matrix}
  \right],$$
  where $\{\beta_i\}$ are $1$ or $\alpha$. But by Lemma \ref{lem1}, the above matrix is invertible. So we only need to look at the remaining sub matrix. Again, we can take out another small block column and row from $\overline{I}$ from each big block, and it is invertible by Lemma \ref{lem1}. Continue this process, we are left with only columns and rows of $I$ in each big block. For all $i,i' \in I, 1 \le i'-i < \frac{r}{2}$ or $i'-i=\frac{r}{2}, i < \frac{r}{2}$, consider row $i'$ in $A^i$ and row $i$ in $A^{i'}$. They are shown in \eqref{eq2}. One can do row operations and keep the invertibility of the matrix, and get
  $$
    \bordermatrix{
        &        & \hspace*{-0.3cm} i & \hspace*{-0.3cm}        & \hspace*{-0.3cm}  i'             &  \hspace*{-0.3cm}       & \hspace*{-0.3cm} i  &    \hspace*{-0.3cm}     & \hspace*{-.5cm}  i'             &     \hspace*{-.5cm}    \cr
     i' \text{ in } A^i& \cdots & \hspace*{-0.3cm}0 & \hspace*{-0.3cm}\cdots & \hspace*{-0.3cm} p_{j_1}^{i-i'} &\hspace*{-0.3cm} \cdots &\hspace*{-0.3cm} 0  & \hspace*{-0.3cm} \cdots & \hspace*{-0.3cm} p_{j_t}^{i-i'} & \hspace*{-0.3cm}\cdots \cr
     i  \text{ in } A^{i'}& \cdots & p_{j_1}^{i'-i} & \hspace*{-0.3cm}\cdots &\hspace*{-0.3cm} 0 & \hspace*{-0.3cm}\cdots & \hspace*{-0.3cm}p_{j_t}^{i'-i} & \hspace*{-0.3cm}\cdots &\hspace*{-0.3cm} 0 & \hspace*{-0.3cm}\cdots}.
  $$
  Proceed this for all $i,i' \in I$, we are left with block diagonal matrix in each big block and the matrix left is of size $t^2 \times t^2$. Taking out the $i_1$-th column and row in each big block, we have the following $t \times t$ sub matrix:
  $$\left[\begin{array}{ccc}
    p_{j_1}^0 &    \cdots & p_{j_t}^0  \\
    p_{j_1}^{i_2-i_1}&  \cdots & p_{j_t}^{i_2-i_1}  \\
    \vdots & & \vdots \\
    p_{j_1}^{i_t-i_1}&  \cdots & p_{j_t}^{i_t-i_1}  \\
  \end{array}
  \right],$$
  which is invertible by Lemma \ref{lem1}. Similarly, we can take out the $i_2$-th column and row, and so on, and each sub matrix is again invertible. Thus, any matrix $A$ is invertible and Construction \ref{cnstr2} is MDS.
\end{IEEEproof}

For example, one can easily check that the code in Figure \ref{fig3} is able to recover the information from any two nodes. Therefore it is an MDS code.

\section{Summary}
\label{sec4}
In this paper, we presented constructions of MDS array codes that achieve the optimal rebuilding ratio $1/r$, where $r$ is the number of redundancy nodes. The new codes are constructed based on our previous construction in \cite{old} and improve the efficiency of the rebuilding access.

Now we mention a couple of open problems. For example, if there are $k-1$ systematic nodes and $r$ parity nodes, then our code has $r^k$ rows. Namely, the code length is limited, are there codes that are longer given the number of rows? For example, when $r=2$, we know an optimal rebuilding ratio construction with $r^k$ rows and $k$ systematic nodes:
$$A_j^0=\left[\begin{matrix}
  I & 0 \\
  p_j & I
\end{matrix}
\right],
A_j^1=\left[\begin{matrix}
  I & p_j \\
  0 & I
\end{matrix}
\right].
$$
Here $A_j^0,A_j^1$ are the matrices that generate the parities, and we can take all $j \in [1,k]$. On the other hand, given $r^k$ rows, it can be proven that any systematic and linear code with optimal ratio has no more than $k+1$ systematic nodes. Thus the proposed code length can be improved by at most $2$ nodes.

Finally, using the code in \cite{old} one is able to rebuild any $e, 1 \le e \le r$, \emph{systematic} erasures with an access ratio of $e/r$. However, it is an open problem to construct a code that can rebuild any $e$ erasures with optimal access.

\section*{Acknowledgment}
We thank  Dimitris Papailiopoulos, Alexandros Dimakis and Viveck Cadambe for the inspiring discussions.


\end{document}